\documentstyle[11pt,epsf]{l-aa}
\def\arcmin              {$^{\prime}$}

\def\arcsec              {$^{\prime\prime}$}

\def\kmsmpc             {km\thinspace s$^{-1}$\thinspace Mpc$^{-1}$}

\begin{document}
\title{Detection of the  lensing galaxy  in HE~1104-1805 \thanks{Based
on observations obtained at ESO, La Silla, Chile}}
\author{F. Courbin\inst{1,2}, C. Lidman\inst{3} and P.  Magain
\inst{1}\thanks{Also   Ma\^{\i}tre  de   Recherches  au  FNRS
(Belgium)}}
\thesaurus{12.07.1;11.17.4;03.13.2}
\offprints{F. Courbin (Li\`ege address)}
\institute{
Institut d'Astrophysique, Universit\'e de Li\`ege,
Avenue de Cointe 5, B--4000 Li\`ege, Belgium
\and
URA 173 CNRS-DAEC, Observatoire de Paris,
F--92195 Meudon Principal C\'edex, France
\and
European Southern Observatory, Casilla 19001, Santiago 19, Chile.}
\date{Submitted ; Accepted}
\maketitle
\markboth{F. Courbin et al.: The lensing galaxy in HE~1104-1805}{}

\begin{abstract}

We report  on deep IR  imaging of the double  quasar HE~1104--1805.  A
new  image  deconvolution technique has  been  applied to  the data in
order   to  optimally combine  the   numerous  frames  obtained.   The
resulting $J$ and $K'$ images allow us to detect and study the lensing
galaxy between  the two lensed  QSO images.  The  near infrared images
not  only confirm the  lensed nature  of  this double quasar, but also
support   the previous redshift estimate   of $z=1.66$ for the lensing
galaxy.  No obvious overdensity    of  galaxies is detected  in    the
immediate region surrounding the lens,  down to limiting magnitudes of
$J=22$ and $K=20$.  The geometry of the system, together with the time
delays expected  for     this lensed quasar,     make  HE~1104--1805 a
remarkable target for future photometric  monitoring programs, for the
study of microlensing and for   the determination of the  cosmological
parameters in the IR and optical domains.


\end{abstract}

\section{Introduction}

It is well established that  gravitationally lensed quasars are unique
natural rulers   for  measuring  the Universe  and  for   deriving the
cosmological parameters  (Refsdal, 1964a,b).  Measuring the time delay
from the images of a lensed QSO can  provide an estimate of the Hubble
parameter $H_{0}$,    independent  of any    other  classical  method.
However, a  good knowledge of  the geometry  of the lensed  system  is
mandatory for the method to be effective (e.g.  Schechter et al, 1997;
Keeton \& Kochanek,  1996;  Courbin et al,  1997).   In spite of  this
crucial requirement and although  the number of  known gravitationally
lensed quasars does not  stop increasing (see for  a review Keeton  \&
Kochanek, 1996), the  precise  geometry of  most lensed  QSOs  remains
poorly known.  In most   cases, even the  matter responsible   for the
lensing, whether it  be  in the form of  a  single galaxy or   several
galaxies,  is  not  detected.  The  high  redshifts of  these galaxies
(hence their faint apparent  magnitudes) and the strong blending  with
the nearby  much brighter QSO  images are the  main reasons  for their
non-detection.

Imaging in the near IR  (1 to 2.5 microns)  has the advantage that the
relative  brightness between the  lensed   QSO and any lensing  galaxy
decreases, making the  galaxy easier to   detect. The disadvantage  is
that the IR sky is considerably brighter. This forces one to take many
images to avoid detector saturation; however, this turns  out to be an
advantage (see Section 3).

This paper presents IR observations of the  quasar HE~1104--1805.  The
strong  similarity between  the  optical spectra obtained  for its two
components   (Wisotzki  et   al, 1993)   makes    HE~1104--1805 a good
gravitational  lens  candidate.   The  high  redshift  of  the  object
($z=2.316$,  Smette   et al, 1995) and    the  relatively wide angular
separation between  the lensed  images  (3.2\arcsec) indicate  that an
large mass is involved in the  lensing potential.  If the deflector is
a  high  redshift galaxy  or a  galaxy  cluster,  deep IR observations
should reveal it.

We used a  recently  developed image deconvolution algorithm  (Magain,
Courbin   \& Sohy,  1997; hereafter  MCS)   to  optimally combine  the
numerous IR  frames and  obtain deep,  sharp images of  HE~1104--1805.
The  present paper describes how  this powerful technique allows us to
study the  immediate environment   of  HE~1104--1805 and   detect  the
lensing galaxy.

\section{Observations-reductions}

The observations took place at the  ESO/MPI 2.2m telescope situated at
La Silla Observatory, Chile,  on the nights of April  14 and 15, 1997.
The  IR camera  IRAC2b   was used  at  the  Cassegrain  focus   of the
telescope.  The detector of IRAC2b is a 256$\times$256 NICMOS~3 HgCdTe
array and the instrument has a variety of optical lenses available for
imaging at different pixel scales.  Lens LB (Lidman, Gredel \& Moneti,
1997) was chosen since it  gives a good  compromise between the  pixel
scale (a small pixel is needed for the  deconvolution) and the size of
the field.  During the observations, PG~1115+080 was also observed.  A
by-product  of these observations is a  more  accurate estimate of the
IRAC2b pixel size  for  images  taken  in lens  LB.   The  new  scale,
0.2762\arcsec /pixel, is based on the precise astrometry of this field
given by  Courbin et al. (1997).  This results in  a  field of view of
71\arcsec.

Numerous   short exposures of  HE~1104--1805   were  obtained in   $J$
($\lambda_c$=1.25 micron)  and  $K'$ ($\lambda_c$=2.15  micron)  under
good meteorological conditions.  The mean seeing was 0\farcs7-0\farcs8
and the sky  was photometric.   We  set the Detector Integration  Time
(DIT) to 20 sec in  $K'$ and 60 sec in  $J$.  Each image taken in $K'$
(resp. $J$) is the average  of  3 (resp.  2)  such integrations.   The
choice of the  DIT is dictated  by detector saturation.  The number of
DITs is dictated  by the frequency at  which the sky intensity varies.
Since the field of HE~1104--1805 is  uncrowded, we took sequences of 9
science exposures, dithered in a semi random manner by 5 to 10\arcsec,
always keeping the object and three PSF stars in the field.

Dome flat-fields were taken in order to correct for the pixel to pixel
sensitivity variations of  the detector.  However, dome flat-fields do
not accurately represent the large scale sensitivity variations of the
array,  in the  $J$ band.   Fortunately, this  variation does not show
strong gradients and has  a maximum amplitude  of 3-4\% over the whole
field. It was modeled   by observing a bright  star  over a grid  of 9
different positions across the array and by fitting a two dimensional,
third order polynomial to the flux of  the star.  The residuals of the
fit were  1\%.  This  fit, which is   commonly called an  illumination
correction, was  multiplied by  the dome   flat to  produce  the final
flat-field containing  both the  low  and  high frequency  sensitivity
variations  of  the array.  After subtraction    of a dark  frame, the
flat-field correction is applied to all scientific frames.

The background is removed from every exposure. It is estimated for any
particular exposure by  averaging the 6 preceding  and the 6 following
exposures.  Thanks to  the  dithering  between exposures,  objects  in
these frames could  be  rejected before the  average  was taken.  This
method allows us to accurately  follow the background which varies  on
the time scale of a few minutes.

Standard stars were observed every  two hours. The standard  deviation
in the zero  points were 0.024 magnitudes in  $J$ and 0.012 magnitudes
in  $K$.  The magnitude and  colours  in this  paper  are in the $JHK$
system as defined by Bessell and Brett (1988).

\section{Image Combining/Deconvolution}

The  frames were combined in  two ways.  First, the standard reduction
and image combination techniques implemented  in the IRAF package were
used in order to average  the frames. The sigma-clipping algorithm was
used  for bad pixel  rejection.  This leads  to two deep  $J$ and $K'$
images over a field  of 1\arcmin.  The total  exposure times were 5040
sec in $J$ and 8100 sec  in $K'$. The  resulting detection limit is 22
in    $J$ and  20   in  $K$   (3$\sigma$, integrated   over  the whole
object). Fig.  1 presents the field in the $J$ band.

\subsection{Image Deconvolution}

\begin{figure}[t]
\begin{center}
\leavevmode 
\epsfxsize= 8.5 cm 
\epsffile{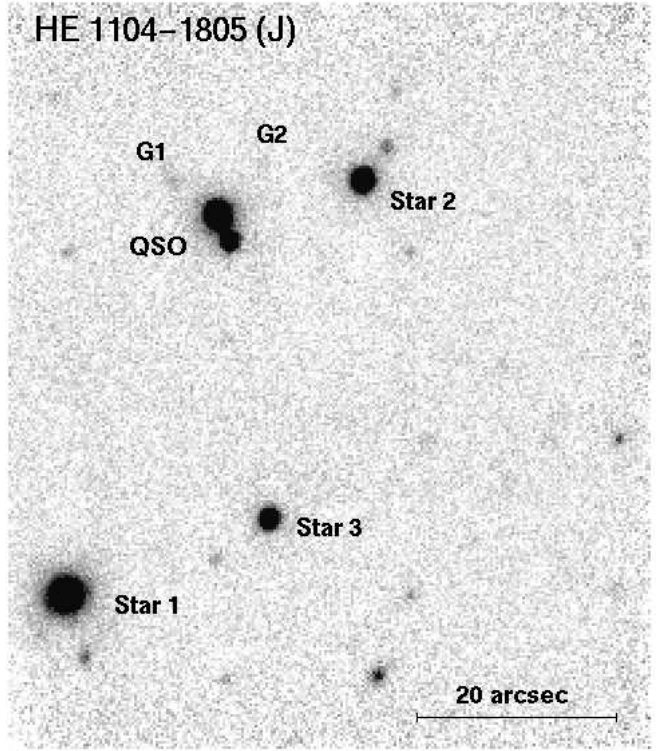}
\caption{A  field   of   one arc-minute around    HE~1104--1805.   The
detection limit on this  deep $J$ band image is   22.  West is to  the
top, North to the left. The PSF  stars are labeled.  No obvious galaxy
overdensity near the QSO pair can be  seen; only galaxies G1 and/or G2
might be involved in the lensing potential.}
\end{center}
\end{figure}
\begin{figure*}[t]
\begin{center}
\leavevmode 
\epsfxsize= 17.5 cm 
\epsffile{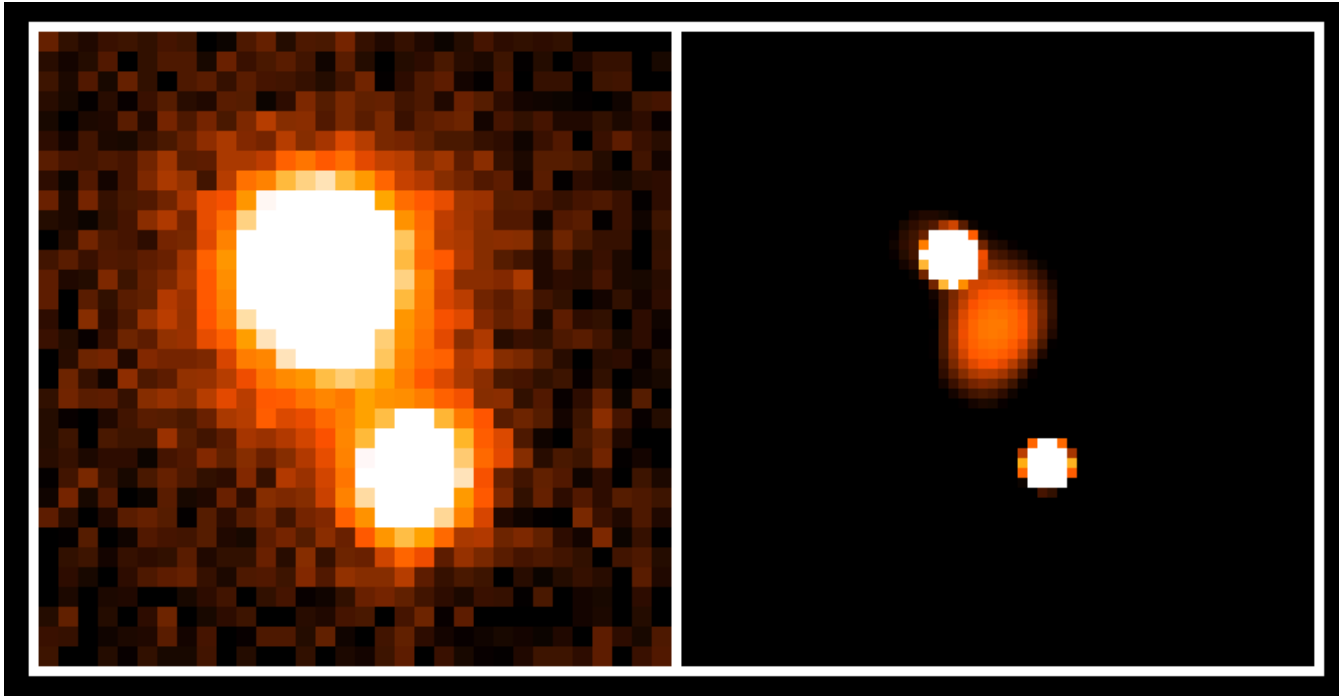}
\caption{{\it Left}: A $J$  band image of HE~1104--1805 obtained  with
2.2m  ESO/MPI  telescope.   The   field is 8.8\arcsec   and  the total
exposure time is 5040 sec.  {\it Right}: Simultaneous deconvolution of
the 6 intermediate   images (see text).    The adopted  pixel  size is
0\farcs1381. The FWHM of the point sources on the deconvolved image is
0\farcs2762.}
\end{center}
\end{figure*}

In order to study the immediate  environment of HE~1104--1805, we used
the  new MCS  deconvolution algorithm   described  in full detail   by
Magain, Courbin \& Sohy (1997).

Deconvolution of an image  by the total observed Point-Spread-Function
(PSF) leads to  the so-called ``deconvolution artifacts'' or ``ringing
effect''    around   the point  sources.      This  results  from  the
deconvolution  algorithm  attempting  to  recover  spatial frequencies
higher  than  the   Nyquist frequency,   thus  violating the  sampling
theorem. Instead, the MCS algorithm uses  a narrower PSF which ensures
that the deconvolved image   will  not violate the  sampling  theorem.
Additionally, the   MCS algorithm takes  advantage   of important {\it
prior knowledge}:  in  the deconvolved  images,  all the point sources
have the   same  (known) shape.    This  allows us   to  decompose the
deconvolved image   into a  sum of  analytical  point  sources plus  a
diffuse  background  which is  smoothed to  the final  resolution {\it
chosen by the  user}.  Most  of the  deconvolution artifacts  are thus
avoided.  This is of particular interest  when one wishes, like in the
present study, to discover faint  objects embedded in the seeing disks
of much brighter point sources.

If  the  deconvolution of a    single image already  yields very  good
results (e.g.  Courbin et al. 1997), the simultaneous deconvolution of
numerous dithered exposures  is even more  efficient (e.g.  Courbin \&
Claeskens, 1997).  In particular, the  MCS code allows the pixel  size
of    the deconvolved   image  to  be  as  small     as desired.  This
over-sampling possibility, already applicable  to the deconvolution of
a single   frame, is of   considerable interest when  dealing with the
spatial information contained in many dithered frames.

Another advantage of the MCS  algorithm is that the  PSF can vary from
frame to frame.  For example, one can combine good quality images with
trailed or even defocused images or, in a  more reasonable way, frames
of differing image  quality and signal-to-noise ratios.  The resulting
frame  is  an optimal combination  of the   whole data  set, with {\it
improved resolution and sampling}.

The seeing in the original IRAC2b images was of  the order of 0\farcs6
for the  very best frames and up  to 1\farcs2 for  the worse  ones, in
both $J$ and $K'$.   We adopted for the  deconvolution a sampling step
of 0\farcs1381, two  times smaller than  the original pixel size. This
allows   us to  reach a  final  resolution  of 0\farcs2762 which still
samples well   the resulting image (2   pixels Full-Width-Half-Maximum
(FWHM)).

Since  the signal-to-noise of individual images  is very low and since
we  had to reject very  numerous   bad pixels,  we first combined  the
images in groups of nine.  Thus, we obtained  6 intermediate images in
$J$ and 12 images   in $K'$.  A  PSF was   derived for each  of  these
images.  In $K'$, only ``Star 1'' is bright enough, - i.e.  comparable
to the QSO's luminosity - to compute an  accurate PSF (See Fig.  1 for
the labeling  of the stars used).  In  $J$, ``Star  1'' shows extended
luminosity so  we used both ``Star 2''  and ``Star 3''.  The resulting
total  exposure time of the co-added  images is different from the one
of  the images  combined  using IRAF  and  the standard   methods.  We
rejected more frames  with bad pixels falling right  on the  object or
the PSF stars.  On the other hand, we included in the different stacks
more images with bad  seeing.  Thus, the  total  exposure time  of the
deconvolved images is 6480s in both $J$ and $K'$.

The    program   requires initial  estimates    for  the positions and
intensities  of the  point  sources in  the field.   This  was done by
choosing the central pixel  of each QSO image.  During  deconvolution,
the centres of the point sources are forced to be the  same in all the
images,  only an image translation  (no  rotation) being allowed.  The
data  are  never aligned or  rebinned;  only the deconvolved model (on
which  the highest spatial  frequencies  are modeled analytically)  is
transformed.  The  intensities of the point  sources can be allowed to
be different in  each   image so that  even  variable  objects may  be
considered in the deconvolution.

The shape of IRAC2b PSF shows significant variations across the field.
In $J$, the   variation is still  acceptable, mainly  because we  used
``Star 2''  and ``Star  3'',  which are closer to   HE~1104--1805 than
``Star 1'', which  is used for the  PSF computation in the  $K'$ band.
It is possible in  our algorithm to let the  PSF depart slightly  from
its  original shape, during the  deconvolution process.  It is in fact
re-determined directly  from the point  sources in  the  field that is
deconvolved.  In the present  case of simultaneous  deconvolution, the
correction  on the  PSFs  is well constrained  by the  numerous images
considered.   The   quality of the PSF   correction  is even better if
numerous stars are present in the field.   With 2 point sources and 12
images in $K'$,  it has been possible to  correct rather well the PSFs
of the 12 images.  The  deconvolution is first performed with variable
PSFs, and then repeated with the corrected PSFs fixed.

\begin{figure}[t]
\begin{center}
\leavevmode 
\epsfxsize= 9.0cm 
\epsffile{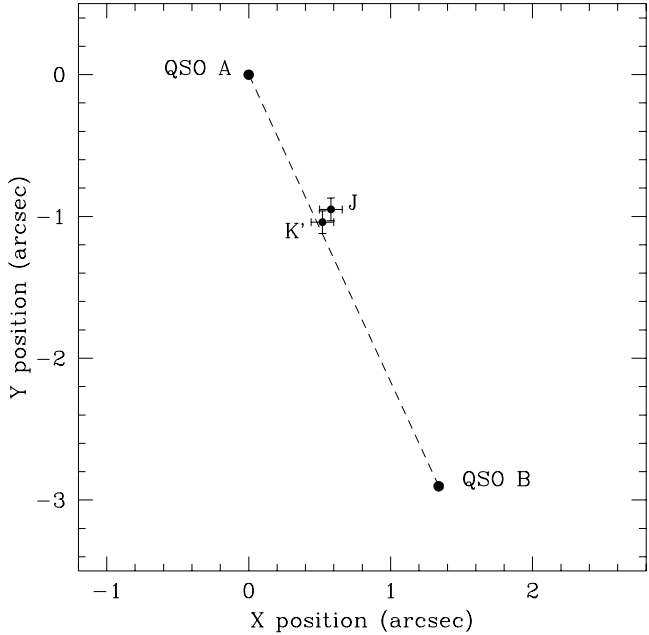}
\caption{Geometry of  HE~1104--1805.   Note the   slight  misalignment
between the 2 QSOs and the lensing galaxy.}
\end{center}
\end{figure}

The background  component of   the  deconvolution is  smoothed on  the
length scale of the  final resolution.  The  weight attributed to  the
smoothing  (see  Magain, Courbin  \& Sohy,  1997   for more detail) is
chosen so that the residual map between each  data frame and the model
image (reconvolved with the PSF) in units of the photon noise, has the
correct statistical distribution, i.e.  is equal to unity all over the
field.  In other words, we chose the  smoothing term by inspecting the
local  residual maps.   This  ensures  that  the deconvolved image  is
compatible with the whole  data set in any region  of any of the  data
frames.

The deconvolution  consists  of a $\chi^{2}$  minimization between the
deconvolved  model image and    {\it the whole  data  set},   using an
algorithm  derived from  the  conjugate  gradient method.  Again,  the
residual maps are used  as a quality check of  the result. We stop the
iteration process  only when  the   residual  maps show   the  correct
statistical  distribution all over  the field  so  that we avoid local
over or under-fitting.

The  program produces the following outputs:  a deconvolved image, the
centre of  the point   sources,  the shifts   between the  images, the
intensities of the point sources for each of the individual frames and
an image  of the deconvolved galaxy,  free of any contamination by the
QSOs.

\subsection{Results}

Figure 2  displays the  result of the  deconvolution  for the $J$ band
images.    Six images were  used to  obtain this  result.  The spatial
resolution is 0\farcs2762, comparable to the resolution reached by the
HST   in the IR domain.   We chose the  same final  resolution for the
simultaneous deconvolution  of 12 $K'$ images.   The lensing galaxy is
clearly detected and  displayed in Fig.  2.   It is also seen in $K'$,
were it is in fact brighter.

\begin{table}[t]
\begin{center}
\caption{Summary of the astrometry (in the  same orientation as Fig. 1
and Fig.  2) and photometry for HE~1104--1805  and the lensing galaxy.
The 1$\sigma$ error bars are also indicated.}
\begin{tabular}{l r r r}
\hline \hline
    & QSO A & QSO B    & Lens \\
\hline  
$J$ & $15.94 \pm 0.06$  &   $17.47 \pm 0.08$  &  $19.01 \pm 0.2$\\
$K$& $14.78 \pm 0.08$  &   $16.13 \pm 0.11$  &  $17.08 \pm 0.2$\\
$J-K$  & $ 1.16 \pm 0.09$  &   $ 1.34 \pm 0.13$  &  $ 1.93 \pm 0.3$\\
\hline
x (\arcsec)&$0.00 \pm 0.03$ &   $+1.34 \pm 0.03$  &  $+0.55 \pm 0.05$\\
y (\arcsec)&$0.00 \pm 0.03$ &   $-2.90 \pm 0.03$  &  $-1.00 \pm 0.05$\\
\hline
\end{tabular}
\end{center}
\end{table}

The images  were  deconvolved several  times,  with  different initial
guesses as to the   position and the intensity   of the QSO pair.   In
Table 1 the relative positions of  the QSOs are tabulated.  The errors
correspond to   the    dispersion in the    different   deconvolutions
(1$\sigma$ error bars).

The photometry of the QSO images  is also given.  The 1$\sigma$ errors
correspond to the dispersion  in the peak intensities  in each of  the
images considered in  the simultaneous deconvolution (6  in $J$, 12 in
$K'$).

The  position of the lensing galaxy  was determined on the deconvolved
background  image by both  Gaussian   fitting and first  order  moment
calculation.  The   results were averaged together  and  taken  as the
position  of the lensing galaxy.   We  estimate the 1$\sigma$ error on
the galaxy  position to be about  0\farcs08 in  both bands. The values
given in Table 1 are the average of the positions  in $J$ and $K'$ and
have an estimated error  of 0\farcs05.  The angular separation between
the lensing galaxy and QSO~A  is 1.14\arcsec$\pm$ 0.06\arcsec and  the
distance between the two QSO images is 3.14\arcsec$\pm$ 0.04\arcsec.

We derived the magnitude of the lensing  galaxy by aperture photometry
on  the deconvolved  background image  to  avoid contamination by  the
QSO's light. A diaphragm of 0\farcs9 diameter was used. Due to the too
low  signal-to-noise  ratio  in  the  lensing  galaxy, we   could  not
determine its shape parameters.

\begin{table}[t]
\begin{center}
\caption{Astrometry and photometry of  galaxies G1 and G2, relative to
QSO A. The astrometry is given in  the same orientation  as Fig. 1 and
Fig.   2. These  values   were  derived from  the   ``un-deconvolved''
images.}
\begin{tabular}{l r r}
\hline \hline
    & G1 & G2 \\
\hline  
$J$   & $20.3 \pm 0.3$  &   $21.4 \pm 0.4$  \\
$K$   & $19.2 \pm 0.3$  &   $19.7 \pm 0.5$  \\
$J-K$ & $ 1.1 \pm 0.4$  &   $ 1.7 \pm 0.6$  \\
\hline
x (\arcsec)&$-4.7 \pm 0.1$ &   $+4.2 \pm 0.1$ \\
y (\arcsec)&$+3.1 \pm 0.1$ &   $+5.0 \pm 0.1$ \\
\hline
\end{tabular}
\end{center}
\end{table}

Figure 3 shows the position of the galaxy, relative to the QSO images.
A slight misalignment between the lens and the  line joining QSO~A and
QSO~B can be seen. It is larger than our error bars and is apparent in
both $J$  and $K'$. In addition,   the PSF's shape   does not show any
significant variations across  the deconvolved field (only 8.8\arcsec)
so  that any geometric  distortion   can be  ruled out. The   observed
misalignment seems therefore real.

No  obvious galaxy overdensity is  detected in the immediate surrounds
of the QSO, although  the detection limit  of 22 in  $J$ and 20 in $K$
would have allowed us  to see any rich  cluster up to $z=2$.  The  two
nearest objects to the double QSO are galaxies $G1$ and $G2$ (see Fig.
1). Table   2 gives  their   position relative   to  QSO A  and  their
photometry,  both derived  on the ``un-deconvolved''  image since they
are outside the field used for the deconvolution.

\begin{figure}[t]
\begin{center}
\leavevmode 
\epsfxsize= 9.0cm 
\epsffile{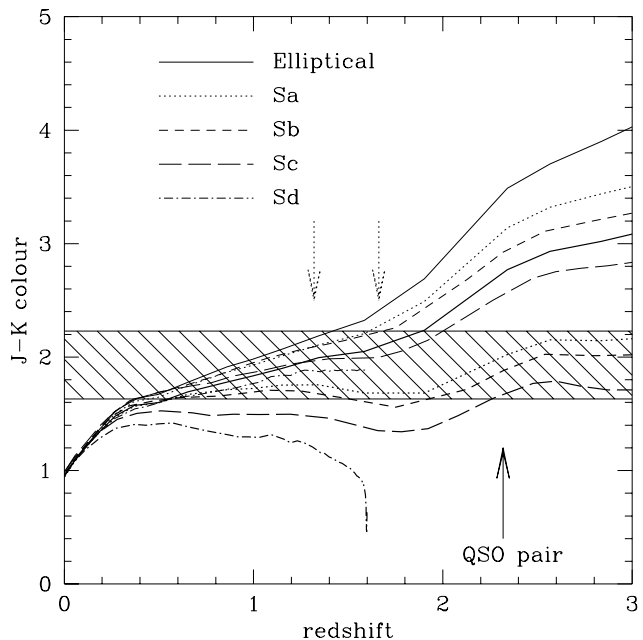}
\caption{$J-K$ as  a function of  redshift for different  galaxy types
(see section 4 for more details).  The shaded  region shows the colour
of the lensing galaxy, error bars included.  The  thick lines show the
models which  take into account   galaxy evolution; the  thin lines do
not.  A  solid  arrow shows the  redshift  of the  QSO  pair.  The two
strongest metal  absorption  line systems, at redshifts  $z=1.320$ and
$z=1.6616$, are marked with dotted arrows.}
\end{center}
\end{figure}

\section{Colour of the deflector}

The redshift of the  lensing galaxy can   be constrained by  the $J-K$
colour.  In   Figure 4, we compare   the colour of  the candidate (the
shaded region) with theoretical  colours corresponding to five  galaxy
types.   These colours  were obtained  from  the  PEGASE  {\it "Projet
d'Etude des  GAlaxies par Synth\`ese  Evolutive"}  atlas (Leitherer et
al. 1996; Fioc and  Rocca-Volmerange 1997).  The five types correspond
to E, Sa, Sb, Sc and  Sd galaxies in a  critical density universe with
$H_0=50$  (Rocca-Volmerange and Fioc  1996).   The theoretical colours
are a function of  redshift.  The thick  lines  include the  effect of
galaxy evolution; the thin lines do not.

\begin{figure}[t]
\begin{center}
\leavevmode 
\epsfxsize= 9.0cm 
\epsffile{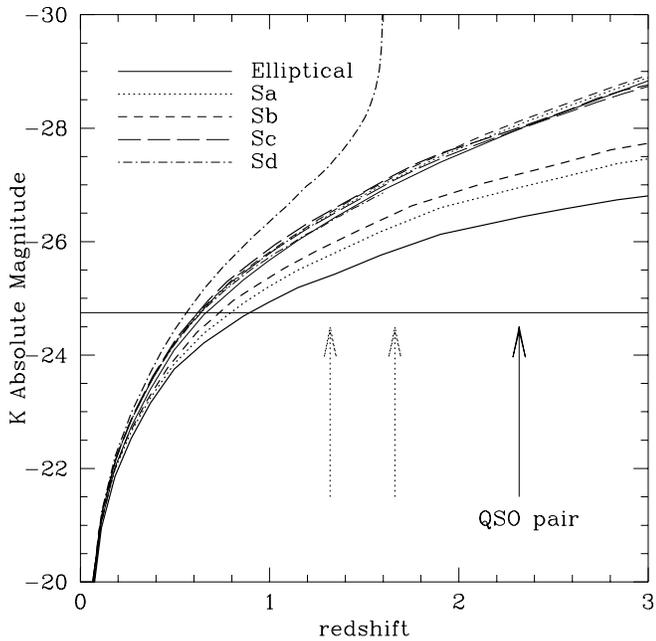}
\caption{Absolute $K$ magnitude as a  function of redshift.  The thick
and thin  curves  have  the same meaning  as  those  in  Fig. 4.   The
redshift of the QSO pair is marked with the solid arrow. The redshifts
of the two strongest metal absorption line systems are marked with the
dotted arrows.  The solid line shows the magnitude of an $M_K^{\star}$
galaxy. (see text)}
\end{center}
\end{figure}

The colour of the candidate constrains the object to  be a galaxy with
a redshift beyond $z=0.4$, probably between z=1 and z=2. The redshifts
of the  two strongest metal absorption  line systems, at $z=1.320$ and
$z=1.6616$ (Smette et al.  1995 and Wisotzki  et al.  1993) are marked
with dotted arrows in Fig. 4 and Fig. 5.

In  Figure 5, we plot the  absolute K-band  magnitude the galaxy would
have    today if it were  at   various redshifts.  The horizontal line
represents the absolute  K-band  magnitude $M_K^{\star}$ of a  typical
large galaxy.  The value  we adopted for  $M_K^{\star}$ is the average
between the value determined by Glazebrook et al.   (1995) and the one
from Mobasher et al.  (1993).  If the galaxy  is at a redshift between
$z=1$ and $z=2$ (as also suggested by Smette et al.  1995), then it is
several times more luminous than an $M_K^{\star}$ galaxy.

\section{Discussion-Conclusions}

The main result of the  present study is  the detection of a red fuzzy
object  located between the   two  components of  HE~1104--1805.  This
result is one more very strong argument in favour of the lensed nature
of this double quasar.

Wisotzki et al (1993) do not detect the lensing  galaxy in $R$ down to
a limiting magnitude of 23-24.   In Fig. 6 the tracks  in the $R-J$ vs
$J-K$ colour-colour diagram of two  galaxy types, an elliptical and an
Sa  galaxy, are plotted.  They are  plotted with and without evolution
and for the redshift range $0  < z < 2.5$.   Also plotted is the range
allowed by the observations in this paper and the optical observations
of Wisotzki et al. (1993).

\begin{figure}[t]
\begin{center}
\leavevmode 
\epsfxsize= 9.0cm 
\epsffile{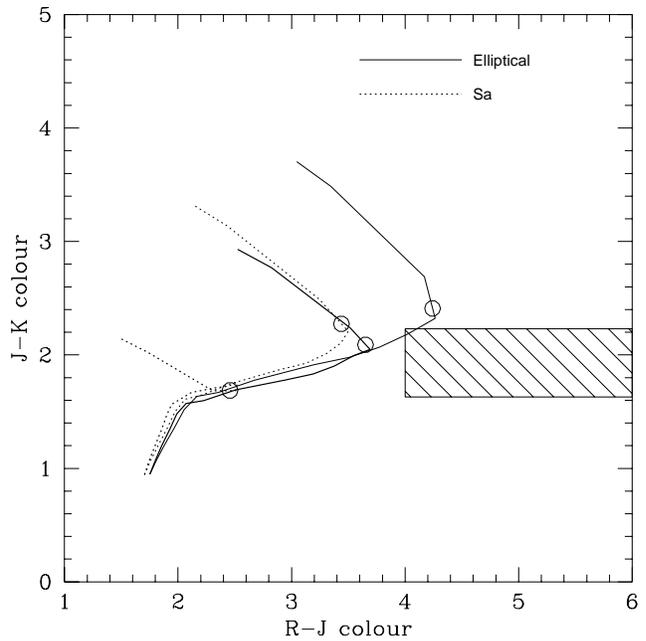}
\caption{A $R-J$ vs. $J-K$ colour-colour plot  showing the tracks from
$z=0$ to $z=2.5$ of two galaxy types,  an elliptical and a spiral. The
thick  lines correspond to the  models  which take into account galaxy
evolution. The location   of a galaxy at $z=1.66$   is  marked by  the
circles.  Also   plotted are the limits derived   from Wisotzki et al.
(1993) and this paper.}
\end{center}
\end{figure}

If we include the  effects of evolution (thick  lines in the Figures),
the IR colours  are compatible with an elliptical  galaxy (as shown by
Fig.  6) between $z=1$ and $z=2$ (Fig.  4 and Fig. 5).  The IR-optical
colours    are less compatible with   this;   however, the expected  R
magnitude  of the lensing galaxy is  $R=22.5$, and  this may have been
difficult to see 1\farcs1 away from the QSO which is 5 to 6 magnitudes
brighter.  In  fact, a preliminary detection of  the lensing galaxy by
Grundahl, Hjorth \& S{\o}rensen  (1995) allowed to measure an $I$-band
magnitude of 20.6, in better agreement with our findings.

One of the two metallic absorption line systems found at $z=1.320$ and
$z=1.6616$  (Smette  et al 1995)   could be   produced by the  lensing
galaxy.  In particular,  the absorption system  at $z=1.6616$ is  seen
almost only  in QSO~A.  Since the  angular  distance from  the lens to
QSO~A is much smaller than to QSO~B  it seems reasonable to think that
the lensing galaxy we detect is more likely to be at $z=1.6616$ rather
than 1.320.

Despite  the  depth  of our IR  images,   which  enables us  to detect
$M_K^{\star}$ galaxies up to a redshift  of the QSO,  we do not detect
any   obvious    overdensity of    galaxies  which  could   contribute
significantly to the  total gravitational  potential involved in  this
system.  However, two faint galaxies (G1 and G2) are detected close to
the line of sight to the QSO.  G1 has a $J-K$ colour  of 1.1, while G2
has  a colour  close  to that of  the  lens galaxy.  These two objects
could constitute an external source  of shear, for example responsible
for the misalignment between the lensing galaxy, QSO~A and QSO~B.

We can  infer   from our  deconvolutions that   QSO~A  is not  exactly
compatible with  a single  point   source.  The  deconvolution  leaves
significant residuals at the location of QSO~A, even  in $J$ where the
PSF is rather stable across the field.   The signal-to-noise ratio and
resolution     of our observations do    not   allow to draw  definite
conclusions,  but we suspect  that image A is  either not single or is
super-imposed on a fuzzy   faint  light distribution.  However,    one
cannot exclude  the simpler (but unlikely)  explanation that the light
and mass centroids of the lensing galaxy do not coincide.

From the geometry of the lensed system, given in Table 1, and assuming
we  see a galaxy  at $z=1.6616$, the  time delay we can expect between
the two images of HE~1104--1805 is of the order of 3.5 years.  We have
assumed that  the lens can be modeled  as a Singular Isothermal Sphere
(SIS) and  that $H_{0}=50$ \kmsmpc.  This  large delay means that  one
measurement every second  week would be   enough to derive  good light
curves.

Assuming that an SIS is appropriate  for the lensing galaxy, we derive
a mass of $7\cdot  10^{11} M_{\odot}$, not  too  far above the  masses
expected for big  elliptical  galaxies.   Fig.   5 supports  that  the
lensing galaxy is bigger than a  ``normal'' galaxy.  If it is actually
at  $z=1.66$   it is one   magnitude brighter  than an $L_{K}^{\star}$
galaxy.

Finally, the magnitude difference between the lensed images is $\Delta
J=1.53\pm  0.1$  and  $\Delta  K=1.35\pm  0.1$, where   the magnitude
difference is   taken  as  $mag(QSO~B)-mag(QSO~A)$.    The   magnitude
difference expected from the SIS model is 0.75 magnitude, but with the
deflector  angularly closer  to the  faint image  than to the brighter
one.

This could  indicate that component  B is  reddened relative  to A, or
that  component A  is   preferentially amplified (e.g.    slight image
splitting) relative  to B and that  this preferential amplification is
more efficient  in  the blue.   The latter  hypothesis is more  likely
since the lens galaxy is  angularly closer to QSO~A  than to QSO~B and
would therefore redden A more than B (assuming the reddening is due to
the lens galaxy).  On the  other hand, the  lensing potential might be
more complex than a SIS (for example elliptical + core), in particular
if G1 and G2 introduce a significant source of shear.

Wisotzki et al (1995) showed that microlensing was acting on QSO~A and
that it was more efficient in the blue than in the red.  The magnitude
difference they observed  in $B$ in  1994 was $\Delta  B=1.85$, larger
than   our present values   in $J$ and  $K$  (although  the quasar has
probably varied between 1994  November and 1997 April).  This suggests
that microlensing is less efficient in the IR than in the visible.  If
the  source quasar is  found to  be variable  in  the IR domain, an IR
photometric   monitoring  of    HE~1104--1805    may  then    minimize
contamination by microlensing events and  allow a better determination
of the time delay than with optical data.

\begin{acknowledgements}
The authors would like to thank J.-P.  Swings, P. Schechter, J. Hjorth
and the referee Y. Mellier  for helpful comments  on the first version
of  this manuscript. This work has  been in part financially supported
by the contract ARC 94/99-178 ``Action  de Recherche Concert\'ee de la
Communaut\'e  Fran\c{c}aise''  (Belgium),   and  P\^ole   d'Attraction
Interuniversitaire P4/05 (SSTC, Belgium).
\end{acknowledgements}

\begin{thebibliography}{}

\bibitem{} Bessell M. S. and Brett J. M. 1988, PASP 100, 1134

\bibitem{} Courbin F., Claeskens J.-F., 1997, ESO messenger 88, p 32

\bibitem{} Courbin F., Magain P. Keeton C.R., et al, 1997, 
A\&A Letters, in press

\bibitem{} Fioc M., and Rocca-Volmerange B., 1997, A\&A, in press

\bibitem{} Glazebrook K., Peacock, J.    A., Miller, L. and   Collins,
C. A. 1995, MNRAS, 275, 169.

\bibitem{}  Grundahl F.,  Hjorth   J.,   S{\o}rensen A.N.,  1995,   in
``Highlights  of  Astronomy'',  Vol. 10 (Kluwer)  Dordrecht,  658-659,
Appenzeler I. (ed.)

\bibitem{} Keeton C.R., Kochanek C.S., 1997, ApJ 487, 42

\bibitem{}  Keeton C.R.,   Kochanek  C.S., 1996,   Proceedings of  the
173$^{rd}$    IAU   Symposium:    ``Astrophysical  Applications     of
Gravitational Lensing'',    Melbourne,  Australia,  eds.    Hewitt and
Kochanek (Kluwer), p 419.

\bibitem{} Lidman C., Gredel R., Moneti A., 1997, ``ESO-IRAC2b user manual''

\bibitem{} Lietherer et al. 1996, PASP 108, 996.

\bibitem{}   Magain P., Courbin  F.,  Sohy  S.,  1997,  ApJ, in press,
preprint astro-ph/9704059

\bibitem{} Mobasher B., Ellis R. S. and Sharples R. M. 1993, MNRAS 263, 560.

\bibitem{} Refsdal S., 1964a, MNRAS 128, 295  

\bibitem{} Refsdal S., 1964b, MNRAS 128, 307

\bibitem{} Rocca-Volmerange B., and Fioc M., 1997, in preparation

\bibitem{} Schechter P.L. et al. 1997, ApJ 475, L85

\bibitem{} Smette  A.,   Robertson J.  G., Shaver  P.  A., Reimers D.,
Wisotzki L., and K\"ohler T. 1995, A\&ASS 113, 199.

\bibitem{} Wisotzki L., K\"ohler T., Kayser  R., Reimers D., 1993, A\&A 278,
L15

\bibitem{} Wisotzki L., K\"ohler T., Ikonomou M., Reimers D., 1995, A\&A 297,
L59

\end{thebibliography}
\end{document}